\documentclass[twocolumn]{aa}
\usepackage{graphicx}
\usepackage{txfonts}
\begin{document}
%
\def\ltsima{$\; \buildrel < \over \sim\;$}
\def\ltsim{\lower.5ex\hbox{\ltsima}}
\def\gtsima{$\; \buildrel > \over\sim \;$}
\def\gtsim{\lower.5ex\hbox{\gtsima}}
\def\ms{$M_{\odot}$ }
\def\msp{$M_{\odot}$}

   \title{Implications of Elemental Abundances in Dwarf Spheroidal Galaxies}

   \author{ T. Tsujimoto}
   
   \offprints{ T. Tsujimoto}

   \institute{National Astronomical Observatory, Mitaka-shi,
                   Tokyo 181-8588, Japan\\
                    \email{taku.tsujimoto@nao.ac.jp}
                   }
    
\date{Received August 30, 2005; accepted September 17, 2005}

\abstract{
Unusual elemental abundance patterns observed for stars belonging to nearby dwarf spheroidal (dSph) galaxies are discussed. Analysis of the [$\alpha$/H] vs.~[Fe/H] diagrams where $\alpha$ represents Mg or average of $\alpha$-elements reveals that Fe from Type Ia supernovae (SNe Ia) does not contribute to the stellar abundances in the dSph galaxies where the member stars exhibit low $\alpha$/Fe ratios except for the most massive dSph galaxy, the Sagitarrius.
The more massive dwarf (irregular) galaxy, the Large Magellanic Cloud, also have an SNe Ia signature in the stellar abundances. These findings suggest that the condition of whether SNe Ia contribute to chemical evolution in dwarf galaxies is likely to depend on the mass scale of galaxies. Unusual Mg abundances in some dSph stars are also found to be the origin of the large scatter in the [Mg/Fe] ratios and responsible for a seemingly decreasing [Mg/Fe] feature with increasing [Fe/H]. In addition, the lack of massive stars in the dSph galaxies does not satisfactorily account for the low-$\alpha$ signature. Considering the assemblage of deficient elements (O, Mg, Si, Ca, Ti, and Zn), all of which are synthesized in pre-SN massive stars and in SN explosions, the low-$\alpha$ signature appears to reflect the heavy-element yields of massive stars with smaller rotation compared to solar neighborhood stars.
 }
 
 \maketitle
%
 
\section{Introduction}

The elemental abundance patterns of stars reflect the combination of nucleosynthesis yields and star formation histories. Now that it is possible to obtain detailed elemental abundances of individual stars beyond the Milky Way thanks to ongoing observations by 8-m class telescopes, nearby galaxies can be used as a laboratory to check the wisdom accumulated to date through the numerous studies on stars in the Milky way (e.g., Wheeler et al. \cite{Wheeler_89}; McWilliam \cite{McWilliam_97}). Such validation will also address the issue of whether the knowledge obtained for our galaxy can be applied to the chemical evolution of other galaxies, and represents a step toward investigation of the distant universe, where information on elemental abundances remains limited and uncertain.

At present, the theoretical understanding of the observed elemental abundance patterns of stars in nearby dwarf spheroidal (dSph) galaxies has come to a standstill. One of the most remarkable abundance features in these galaxies revealed by accumulated observational results ({Shetrone et al. \cite{Shetrone_0103}; Bonifacio et al. \cite{Bonifacio_04}; Geisler et al. \cite{Geisler_05}; Monaco et al. \cite{Monaco_05}) is the deficiency of $\alpha$-elements in comparison with Fe. At first glance, knowledge derived for the solar neighborhood provides a reasonable explanation for the origin of the observed low-$\alpha$ signature as being due to the additional supply of Fe from Type Ia Supernovae (SNe Ia). Based on this supposition, theoretical models of evolutionary change in [$\alpha$/Fe] against [Fe/H] in dSph galaxies have been proposed (Ikuta \& Arimoto \cite{Ikuta_02}; Lanfranchi \& Matteucci \cite{Lanfranchi_04}; Robertson et al. \cite{Robertson_05}). However, there is no compelling evidence for the contribution of SNe Ia to other elemental ratios such as [Mn/Fe] or [$n$-capture/Fe] (see Tsujimoto \& Shigeyama \cite{Tsujimoto_02}). The dSph stars thus exhibit unusual elemental abundance patterns that cannot be explained by current knowledge (see also Shigeyama \& Tsujimoto \cite{Shigeyama_03}).

The chemical evolution of stars is often discussed in terms of the relative abundance ratios between heavy elements, against conventionally [Fe/H]. The relative abundance of an element X with respect to iron ([X/Fe]) is a typical example. In most cases, the [X/Fe] values are confined to much less than 1~dex. Due to the small range of the distributed values, intrinsic scatter partly due to observational error often obscures the fundamental abundance trend. The alternative diagram of two ratios with respect to hydrogen, such as [X/H] vs.~[Fe/H], allows the overall abundance features in the range over $\sim$2~dex to be resolved more clearly. 

  \begin{figure*}
  \vspace{-4.5cm}
   \centering
  \includegraphics[angle=-90,width=\textwidth]{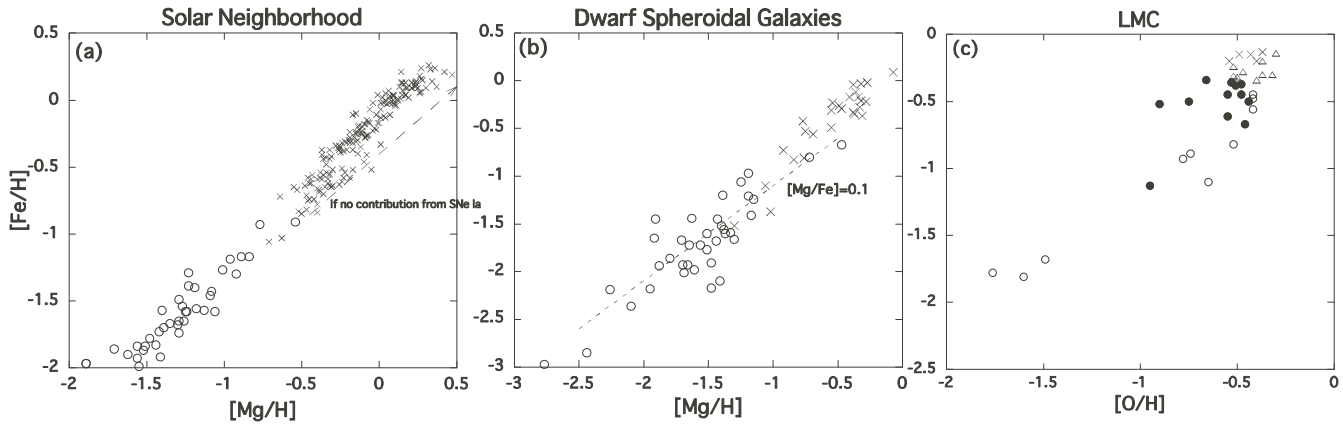}
  \vspace{-1.5cm}
       \caption{(a) Fe and Mg abundances for solar neighborhood stars (open circles; Hanson et al. \cite{Hanson_98}, crosses; Edvardsson et al. \cite{Edvardsson_93}). Dashed arrow denotes the evolutionary path expected without contribution from SNe Ia.  (b) Fe and Mg abundances for dSph stars, showing Draco, Ursa Minor, Sextan, Sculptor, Carina, Fornax, and Leo I (open circles) (Shetrone et al. \cite{Shetrone_0103}; Geisler et al. \cite{Geisler_05}), and Sagittarius (crosses) (Bonifacio et al. \cite{Bonifacio_04}; Monaco et al. \cite{Monaco_05}). Dotted line denotes the correlation [Mg/Fe]~$=0.1$. (c) Fe and O abundances for LMC stars, showing observed data from Smith et al. (\cite{Smith_02})(filled circles), Hill et al. (\cite{Hill_00})(open circles), Hill et al. (\cite{Hill_95})(triangles), and Korn et al. (\cite{Korn_02})(crosses).       
              }
         \label{}
   \end{figure*}

The evolution of [Fe/H] is examined in this study as a function of [$\alpha$/H] in the dSph galaxies, and the results are compared with the data for the Milky Way and the Large Magellanic Cloud (LMC). The LMC is another nearby galaxy that hosts a number of red giants, which also exhibit lower [O/Fe] ratios than stars in the solar neighborhood at the same [Fe/H] (Smith et al. \cite{Smith_02}). This analysis reveals a clear contribution from SNe Ia in the LMC stars as well as stars belonging to the Sagittarius (Sgr) galaxy, but not for stars in other smaller mass dSph galaxies. This is reinforced by analysis of the Mn abundances. 

The origin of the low $\alpha$/Fe ratios for the dSph stars is then discussed. The lack of massive stars of greater than $\sim 20$\ms in the dSph galaxies is proposed as a potential reason for the low $\alpha$/Fe ratios, as the combination of theoretical nucleosynthesis yields for $\alpha$-elements and the Fe mass inferred from SN-light curve analyses predicts that an ensemble of exclusively low-mass Type-II SNe (SNe II) results in low [$\alpha$/Fe]. However, it is shown through abundance studies based on $n$-capture elements and Zn that this possibility is unlikely.
 
The discussion thus reduces to the features of nucleosynthesis in SNe II. The dSph stars exhibit deficiencies in the abundances of not only $\alpha$-elements (O to Ti) but also Zn. Theoretical nucleosynthesis calculations have revealed that some of the depletions arise in pre-SN massive stars, while other deficiencies reflect modification or synthesis in the final SN II explosions (Pagel \cite{Pagel_97}, and references therein). This suggests that a mechanism modifying all stages in the life of massive stars and the corresponding SN yields might be required. This scenario is reminiscent of stellar rotation, which may change both the onion-skin structure during stellar evolution ({Heger et al. \cite{Heger_00}; Heger \& Langer \cite{Heger_00}, Hirschi et al. \cite{Hirschi_04}) and the form of the eventual SN explosions (e.g., Fryer \& Warren \cite{Fryer_04}). It can thus be speculated that the rotation is affected by the environment of stars at birth in such a way that dSph stars develop smaller rotation than in the solar neighborhood stars. 

\section{[Fe/H] vs.~[$\alpha$/H] diagrams}

This analysis starts with an investigation of whether the low $\alpha$/Fe ratios in the dSph stars are promoted by contributions from SNe Ia. If the Fe supply from SNe Ia is added to interstellar matter that has been already enriched by SNe II, the slope of [Fe/H] against [Mg/H] will steepen from the onset of SNe Ia. Namely, after SNe Ia start to contribute, the increase in [Fe/H] becomes larger than that in [Mg/H], while it holds [Mg/Fe]=const until the occurrence of SNe Ia. A good reference for the resolution of such a tendency is undoubtedly the relation for the solar neighborhood, where SNe Ia have been confirmed to contribute to chemical evolution (e.g., Matteucci \& Greggio \cite{Matteucci_86}). Figure Ia shows the [Fe/H] vs.~[Mg/H] diagram for the solar neighborhood stars. As expected, the slope changes at [Fe/H]$\sim -1$ and the bending feature is seen for disk stars, confirming the additional supply of Fe from SNe Ia, although a small fraction of stars belong to thick disk and are distributed on the extension of a trend of halo stars at [Fe/H]\gtsim $-1$ (Feltzing et al. \cite{Feltzing_03}).

   \begin{figure*}
\vspace{-3.8cm}
   \centering
  \includegraphics[angle=-90,width=\textwidth]{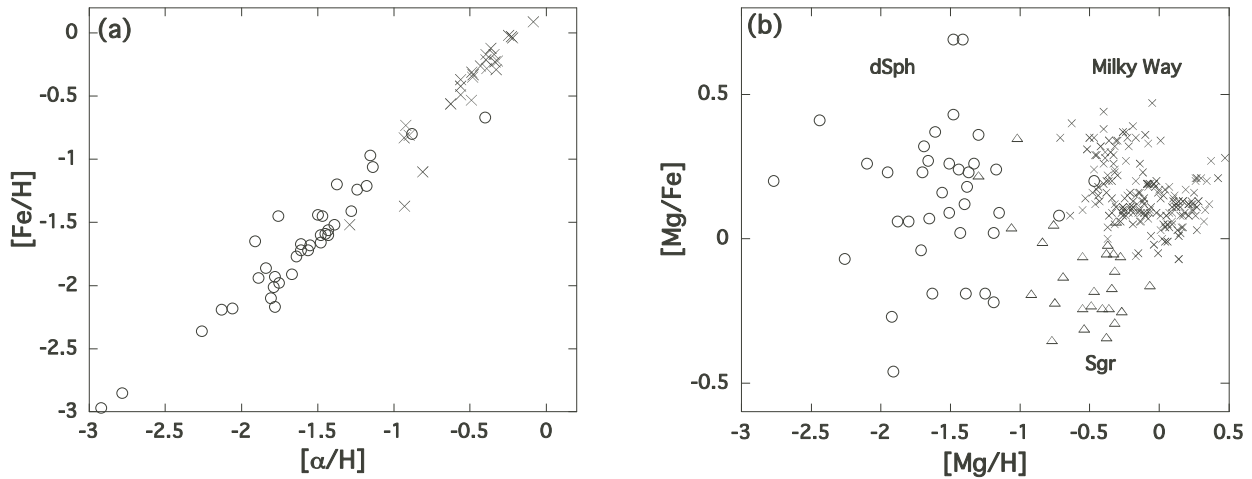}
  \vspace{-1.5cm}
       \caption{(a) Fe and $\alpha$ abundances for dSph stars, showing Draco, Ursa Minor, Sextan, Sculptor, Carina, Fornax, and Leo I (open circles) (Shetrone et al. \cite{Shetrone_0103}; Geisler et al. \cite{Geisler_05}), and Sagittarius (crosses) (Bonifacio et al. \cite{Bonifacio_04}; Monaco et al. \cite{Monaco_05}). Here $\alpha$ is an average of Mg, Ca, and Ti, but  Si is used for the $\alpha$ estimate in Bonifacio et al. (\cite{Bonifacio_04}) due to a lack of Ti data. (b) [Mg/Fe] as a function of [Mg/H] for dSph stars (open circles; Shetrone et al. \cite{Shetrone_0103}, Geisler et al. \cite{Geisler_05}), Sgr stars (open triangles; Bonifacio et al.\cite{Bonifacio_04}, Monaco et al. \cite{Monaco_05}), and disk stars in the Milky Way (crosses; Edvardsson et al. \cite{Edvardsson_93}).
              }
         \label{}
   \end{figure*}

In contrast to the Milky Way, there appears to be essentially no change in the slope of the relation between [Fe/H] and [Mg/H] for the seven dSph galaxies, i.e., Draco, Ursa Minor, Sextan, Sculptor, Carina, Fornax, and Leo I (open circles; Fig.~1b). The bending feature as an evidence of SNe Ia appearance is not seen over the range of $-3$\ltsim [Fe/H] \ltsim $-0.5$. The observed data are distributed along the line [Mg/Fe]~$=0.1$, i.e., $\Delta$[Mg/H]/$\Delta$[Fe/H]~$=1$, indicative of no Fe supply from SNe Ia. Although  the scatter about the line [Mg/Fe]~$=0.1$ is prominent and it renders the contribution of SNe Ia to chemical evolution difficult to judge from this analysis, the absence of such contributions is supported by the Mn abundances, as discussed in the next section. On the other hand, the stars belonging to the Sgr galaxy (crosses; Fig.~1b) appear to exhibit an elemental feature containing the signature of Fe input from SNe Ia, with some clear similarities to the solar neighborhood stars in Figure 1a. It may imply that the far massive Sgr galaxy compared with other dSph galaxies is likely to possess a gravitational potential well that is deep enough for SNe Ia to contribute to chemical evolution. However, a feature common to all the dSph stars is the low Mg/Fe ratio without a clear contribution from SNe Ia. 

The same conclusions for these dSph galaxies can be also deduced from the [Fe/H] vs.~[$\alpha$/H] diagrams where $\alpha$=$\frac{1}{3}$(Mg+Ca+Ti) (due to a lack of Ti data, Si is used for the $\alpha$ estimate in Bonifacio \cite{Bonifacio_04}), as shown in Figure 2a. These $\alpha$-elements have all similar behaviors in the [$\alpha$/Fe] vs.~[Fe/H] diagrams for solar neighborhood stars, and therefore the average ratio of these elements to Fe is also a good indicator of the relative contributions of  SNe Ia and SNe II, although each $\alpha$-element  has a different production process in SNe II, as will be stated in  \S 4.2. Furthermore, Figure 2b offers another view on this subject. The seemingly decreasing [Mg/Fe] features with the increasing metallicity [Fe/H] for dSph stars as well as disk stars in the Milky Way (e.g., Shetrone et al. \cite{Shetrone_0103}; Bonifacio et al. \cite{Bonifacio_04}; Edvardsson et al. \cite{Edvardsson_93}) should be much the same as in the alternative case with the metallicity [Mg/H]  as long as these features are the end results of SNe Ia contribution. Despite a narrower metallicity range given by [Mg/H] compared with [Fe/H] which makes abundance trends of [Mg/Fe] unclear, the decreasing [Mg/Fe] features are evidently seen for disk stars and Sgr stars. In contrast, this is not the case for stars in seven dSph galaxies, showing no sign of SNe Ia contribution to the stellar abundances.

The absence of the contribution from SNe Ia in dSph galaxies except for the Sgr galaxy means that both $\alpha$-elements and Fe are synthesized only in SNe II and thereby their relative evolutions are not affected by variations in star formation history. In other words, the differential chemical enrichment between SNe II and SNe Ia caused by the difference in the abundance patterns of heavy elements ejected through SN explosions as well as the evolutionary timescale of the progenitors will make  various stellar abundance patterns reflecting variations in star formation history of galaxies. One good example demonstrating it might be the LMC as shown in Figure 1c. The LMC stars exhibit an increase in [Fe/H] against a considerably lower $\Delta$[O/H] for [Fe/H]\gtsim $-1$ ([O/H] is used for the LMC due to a lack of observation data). In some stars, the O and Fe abundances appear to decrease, resulting in a bending feature pointing to an opposite direction to the cases of disk stars and Sgr stars. This somewhat complicated abundance behavior is interpreted as representing the end result of Fe supply from SNe Ia combined with a star formation history that differs from that in the solar neighborhood, suggesting the importance of galactic wind in a small galaxy (Venn et al. \cite{Venn_04}) and/or a gap in star formation (Gilmore \& Wyse \cite{Gilmore_91}; Tsujimoto et al. \cite{Tsujimoto_95}). Considering the masses of dwarf galaxies that exhibit an SNe Ia signature in stellar abundances, i.e., the LMC ($5.3\times 10^9$\msp ; Alves \& Nelson \cite{Alves_00}) and the Sgr galaxy ($2-5\times 10^8$\msp ; Law et al. \cite{Law_05}), in comparison with other dSph galaxies with a mass in a range of several $10^6-10^7$ \msp (Mateo \cite{Mateo_98}), the condition of whether SNe Ia contribute to chemical evolution or not is likely to depend on the mass scale of galaxies.

   \begin{figure*}
   \vspace{-4.5cm}
   \centering
  \includegraphics[angle=-90,width=\textwidth]{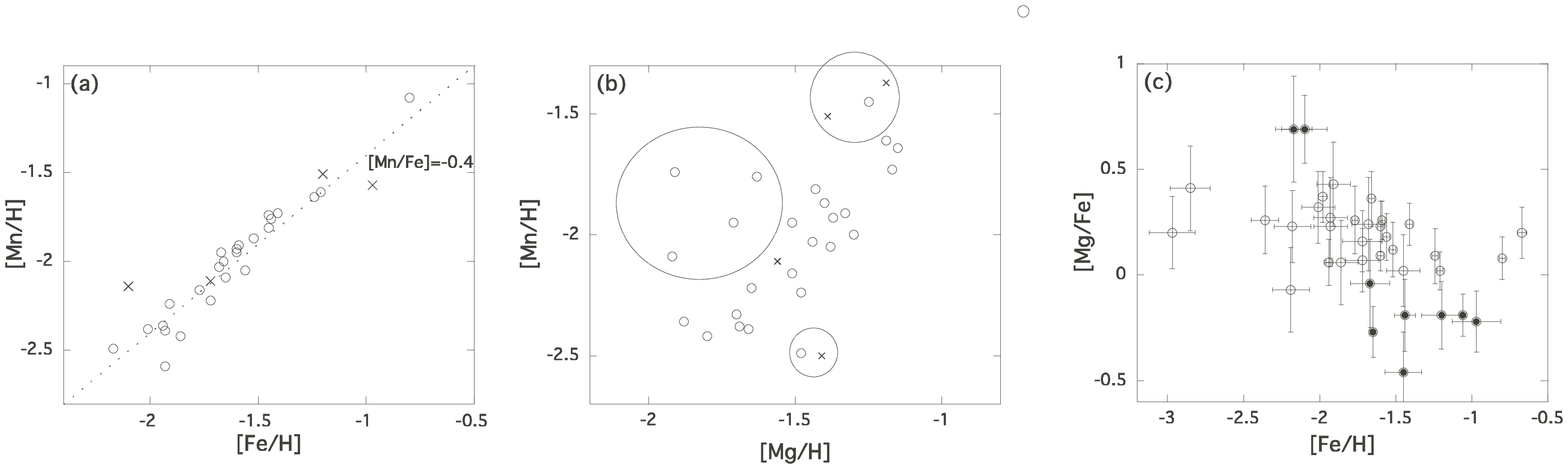}
  \vspace{-2.5cm}
       \caption{(a) Mn and Fe abundances for dSph stars (open circles; Shetrone et al. \cite{Shetrone_0103}, crosses; Geisler et al. \cite{Geisler_05}), including those for globular cluster Palomar 12 which is suggested to originate from the Sgr galaxy (Cohen \cite{Cohen_04}). Dotted line denotes the correlation [Mn/Fe]~$=-0.4$. (b) Mn and Mg abundances for dSph stars (unusual Mg abundances are outlined). The same symbols as for (a). For three stars which largely deviate from the line [Mn/Fe]=$-0.4$ in (a), their Mn abundances are corrected to have [Mn/Fe]=$-0.4$. (c) [Mg/Fe] as a function of [Fe/H] with error bars taken from the literatures. Filled circles denote the stars highlighted in (b). Note that no Mn abundance data is available for some stars of the denoted by open circles. 
              }
         \label{}
   \end{figure*}

In fact, it has been considered that the dSph galaxies have complex star formation histories, many of which have a prolonged star formation over several Gyrs,  from a viewpoint of their color-magnitude diagrams (Mateo \cite{Mateo_98}; Tolstoy et al. \cite{Tolstoy_03}; Venn et al. \cite{Venn_04}; Dolphin et al. \cite{Dolphin_05}).  Such an inferred large span of  ages (see also Grebel \& Gallagher \cite{Grebel_04}) may favor the possibility of the contribution of SNe Ia in the dSph galaxies. However, there is a possibility that the ejecta of SNe Ia easily escaped from the galaxy due to an inefficient cooling by a low density of gas, resulting in little/no contribution of SNe Ia to chemical evolution in spite of a prolonged star formation. As another possible cause, dSph  galaxies were formed through a prolonged accretion over Gyrs of small satellites, in each of which star formation continued for a period of several hundred millions. In any event, a large range of stellar ages does not necessarily require an imprint of SNe Ia on the stellar abundances.

It should be noted that the spectroscopy of large numbers of stars in the Sculptor galaxy has been recently  performed (Tolstoy et al. \cite{Tolstoy_04}) and an initial result on the $\alpha$-abundances (average of Ca, Mg, and Ti) in comparison with Fe has been reported (Tolstoy \cite{Tolstoy_05}). This apparent  decreasing [$\alpha$/Fe] feature with increasing [Fe/H] may recall the SNe Ia contribution again and therefore the acquaintance of abundance data for each element including $\alpha$-elements will be indeed required for further analyses to be pursued. However,  the [Ba/Fe] ratios for these stars coincide with those of halos stars in the Milky Way at the same [Fe/H] (Venn \& Hill \cite{Venn_05}). This coincidence is at odds with an SNe Ia signature in the stellar abundances since the Fe supply from SNe Ia combined with the production of Ba synthesized through $s$-process that is expected to be similar at the same metallicity would lead to the lower [Ba/Fe] ratios than the observed data of halo stars. This result is compatible with the previous reports (Shetrone et al. \cite{Shetrone_0103}), and has been one of the characteristic abundance features that dSph stars exhibit, as mentioned in Introduction.

\section{Mn Tracer in the dSph Galaxies}

The Mn abundance is another important indicator to distinguish products of SNe Ia from those of SNe II, 
whereas other Fe-group elements do not possess such a property. As shown in Figure 3a, the Mn abundance of the dSph galaxies exhibits a strong correlation with the Fe abundance, forming a line of [Mn/Fe]$\sim -0.4$. For the Sgr galaxy,  the data for  globular cluster Palomar 12 ([Fe/H]=$-0.8$) which is suggested to originate from this galaxy is denoted (Cohen \cite{Cohen_04}). The detailed [Mn/Fe] feature for the Sgr galaxy is shown in McWilliam \& Smecker-Hane (\cite{McWilliam_03}), revealing that the Mn/Fe ratios have an SNe Ia signature in the metal-rich regime, which is consistent with the analysis of $\alpha$-elements in \S 2. The Mn-Fe correlation in Figure 3a implies that SNe Ia have not contributed to the Mn and Fe abundances since the [Mn/Fe] ratio would become greater than $-0.4$ if SNe Ia contributed to Mn and Fe in these low-$\alpha$ stars (Shetrone et al. \cite{Shetrone_03};  Shigeyama \& Tsujimoto \cite{Shigeyama_03}). Furthermore, the strength of this correlation can be attributed to the monotonic increase in both Fe and Mn with time, representing good "nucleosynthesis clocks" in the dSph galaxies. 

In contrast, the Mn abundance does not correlate with the Mg abundance (Fig.~3b). Since the Mg abundance should also increase monotonically with time similar to Fe or Mn, irrespective of variations in star formation history, anomalous Mg abundances for some stars are considered to erase such increasing trends. The observed data that deviates substantially from the abundance (Mg)-time (Mn) correlation is indicated in the figure, and the same stars can be seen to exhibit unusual [Mg/Fe] values with a seeming tendency to decrease with increasing [Fe/H] (filled circles in Figure 3c). The less tight correlation between Mn and Mg abundances as a result of excluding the data with unusual Mg abundances may result partially from the relatively large error $\sim 0.2$ dex for the Mg abundances, whereas a typical error of $\sim 0.1$ dex is for the Fe and Mn abundances. In any event, anomalousness of the Mg abundances for the several stars far exceeds measurement errors. Therefore 
it is concluded that unusual Mg abundances in some dSph stars are the origin of the large scatter in the [Mg/Fe] ratios and responsible for a seemingly decreasing [Mg/Fe] feature with increasing [Fe/H].

\section{The Origin of Low-$\alpha$ in dSph Galaxies}

\subsection{ Truncated Initial Mass Function}

Theoretical nucleosynthesis models predict the synthesized mass of Mg in SN II progenitors to a certain extent, whereas the predicted mass of Fe depends on the mass cut chosen for the stellar core and is thus far less certain (e.g., Thielemann et al. \cite{Thielemann_90}). Therefore, the [Mg/Fe] ratio as a function of the progenitor mass $M_{\rm ms}$ cannot be determined purely from a theoretical viewpoint. Here, two approaches supplemented by observational data are employed to derive this relationship. The first approach is based on the abundance patterns of very metal-poor stars, for which recent studies have revealed that stars with elemental abundances of $-4.0$ \ltsim  [Fe/H]  \ltsim $-2.5$ may have inherited the abundance pattern of the ejecta of the preceding SNe (Audouze \& Silk \cite{Audouze_95}; Shigeyama \& Tsujimoto \cite{Shigeyama_98} or the products of the mixing of matter from bursts of the early SNe (Cayrel et al. \cite{Cayrel_04}). Following these hypotheses, observational data (McWilliam et al. \cite{McWilliam_95}; Cayrel et al. \cite{Cayrel_04}) indicate no signs of SNe II with [Mg/Fe]$\sim 0-0.1$. 

Alternatively, on the basis of Fe masses estimated from SN light-curve analyses, the Fe mass increases with increasing $M_{\rm ms}$ for $M_{\rm ms}$ \gtsim 20 \msp, yet remains nearly constant for $M_{\rm ms}$ \ltsim 20 \ms (Shigeyama \& Tsujimoto \cite{Shigeyama_98}; Maeda \& Nomoto \cite{Maeda_03}). As the synthesized Mg mass increases with increasing $M_{\rm ms}$, the breakdown of the linear relationship between the synthesized Fe mass and $M_{\rm ms}$ gives rise to low Mg/Fe ratios for low-mass SNe II. For example, SNe II with $M_{\rm ms}$=13, 15 \ms give [Mg/Fe]=$-0.6$, 0, respectively~(Umeda \& Nomoto \cite{Umeda_02}). If this is the case, the lack of SNe II with $M$ \gtsim $20$ \ms is consistent with the low $\alpha$/Fe ratios for dSph stars (Venn et al. \cite{Venn_04}). It should be noted, however, that the Fe mass estimate by this method has been applied to only a small number of SNe II. In addition, the estimation involves uncertainties due to a dependence on the distance to the galaxies hosting the SNe and uncertain prediction of the progenitor mass (e.g., Van Dyk et al. \cite{Van Dyk_02}). In any event, such a truncated initial mass function (IMF) in the dSph galaxies is at odds with other features as follows.

\vspace{0.2cm}
\noindent 1. {\sl Similarity of $n$-capture/Fe ratios}

The two neutron capture processes, $s$- and $r$-processes,  operate in different astrophysical sites according to different physical conditions; the $s$-process occurs during mixing episodes in low-mass asymptotic giant branch (AGB) stars, whereas the $r$-process may occur during SN II explosions of progenitors within a narrow mass range such as $8-10$\ms (e.g., Travaglio et al. \cite{Travaglio_99}; Ishimaru \& Wanajo \cite{Ishimaru_99}) or $20-25$\ms (Tsujimoto et al. \cite{Tsujimoto_00}). If SNe with $8-10$\ms are assumed for the production site of $r$-process, the truncation of massive stars above $\sim 20$\ms will not affect any of the yields of the $n$-capture elements, while the yields of $\alpha$- and Fe-group elements integrated over all SNe II will certainly decrease. Taking the Salpeter IMF and setting the truncation at $M=15$~\msp, the Fe production rate per generation of stars is reduced by nearly half, resulting in a $\sim 0.3$~dex offset in [$n$-capture/Fe] between dSph stars and solar neighborhood stars. However, the observed data do not support such differences, and instead suggest similar ratios (Shetrone et al. \cite{Shetrone_0103}; Venn et al. \cite{Venn_04}; Venn \& Hill \cite{Venn_05}). Similarly, the choice of $20-25$\ms SNe for the $r$-process site is obviously incompatible with the observation since the production of $r$-process elements hardly occurred in the dSph galaxies with the truncated IMF. In fact, the large uncertainty in the production sites of the $s$- and $r$-processes, including the potential production of $r$-process elements through merging neutron stars and  the operation of the weak $s$-process in massive stars, make discussion on the above offset rather complicated. However,  a difference in a certain degree in the [$n$-capture/Fe] ratios between dSph stars and solar neighborhood stars is unavoidable, given the truncated IMF. 

\vspace{0.2cm}
\noindent 2. {\sl Low Zn/Fe ratio}

Zinc is also deficient in dSph galaxies (Shetrone et al. \cite{Shetrone_0103}; Sbordone \cite{Sbordone_05}). The [Zn/Fe] ratios for solar neighborhood stars essentially resides above the solar ratio (e.g., Cayrel et al. \cite{Cayrel_04}; Nissen et al. \cite{Nissen_04}), whereas dSph stars exhibit [Zn/Fe]$\sim -0.3$ on average with large scatter. Although there is substantial ambiguity regarding the site of Zn production (e.g., Chen et al. \cite{Chen_04}), the high [Zn/Fe] ratios of very metal-poor stars (Cayrel et al. \cite{Cayrel_04}) strongly suggest that SNe II are prominent sites. It has been proposed that effective Zn production can be realized in SNe with large energies (i.e., hypernovae)~(Umeda \& Nomoto \cite{Umeda_02}). Their calculations reveal that low-mass SNe with low [$\alpha$/Fe] ratios produce high [Zn/Fe] ratios of $\sim 0.1$--$0.6$, depending on the explosion energy. This result is distinctly incompatible with the formation of low [Zn/Fe] ratios due to the truncated IMF. The high [Zn/Fe] ratios in low-mass SNe are in fact due to the relatively high Fe production ($\sim 0.07$~\msp) assumed in theoretical models, which requires a large Si-burning layer where Zn is produced. In other words, low [$\alpha$/Fe] ratios necessarily yield high [Zn/Fe] ratios in current models.

\subsection{Stellar Rotation}

The elements with clearly deficient abundances with respect to Fe in dSph stars, compared to halo stars in the solar neighborhood, are O, Na, Mg, Al, Si, Ca, Ti, and Zn. Here, Na and Al are secondary elements, the production of which is limited by the small amount of seed heavy elements (Ne and Mg) contained in the initial chemical composition of the SN progenitor star (see also Venn et al. \cite{Venn_04}). The mechanism modifying the SN yields of the remaining elements ($\alpha$-elements and Zn) is postulated as follows. The elements O and Mg are produced as a result of hydrostatic burning in massive stars, and SN explosions merely expel these elements without affecting their yields. In contrast, the yields of Si and Ca are modified by the final SN explosions, with changes beginning from the formation of the onion-skin structure. Titanium and Zn are also produced in explosive nucleosynthesis. Therefore, the process causing certain elements to be produced less effectively in dSph stars compared to solar neighborhood stars must be active in both pre-explosive massive stars and during SN explosions.

   \begin{figure}
   \vspace{-1.5cm}
\hspace{-1.7cm}
  \includegraphics[angle=-90,width=12cm]{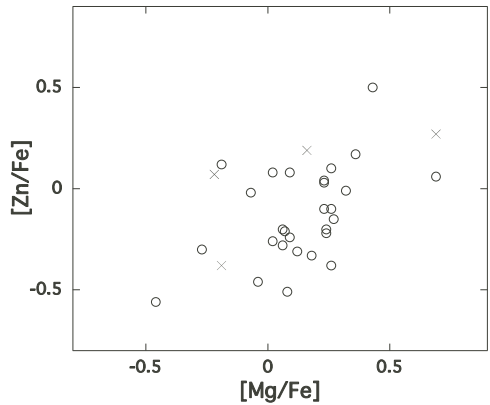}
  \vspace{-0.5cm}
       \caption{[Zn/Fe] as a function of [Mg/Fe] for dSph stars (open circles; Shetrone et al. \cite{Shetrone_0103}, crosses; Geisler et al. \cite{Geisler_05}).
              }
         \label{}
   \end{figure}

Stellar rotation has the potential to produce the observed variety in the yields of all these elements. The effects of rotation on the evolution of pre-SN massive stars have been discussed in detail (Hirschi et al.  \cite{Hirschi_04}). Rotation-induced mixing, by which the products of the burning in the core are mixed into the stellar envelope and new fuel is supplied to the convectively burning stellar core, has been shown to have a marked impact on the internal structure, even altering the deep interior of the core (e.g., mass of the Si core). The calculations by Hirschi et al. (\cite{Hirschi_05}) suggest that the yields of C and O are enhanced by stellar rotation. The Mg yield in a model with $M_{\rm ms}=12$~\ms also increases substantially. Although these results are for stars with solar metallicity, Maeder \& Meynet (\cite{Maeder_05}) reported that for stars with low metallicity, the effects of rotational mixing are also significant, and thus the yields of $\alpha$-elements are enhanced by stellar rotation. Additionally, rotation has been suggested to promote deformed or asymmetric SN explosions, which are likely to involve jet-like explosions (e.g., Yamada \& Sato \cite{Yamada_94}). Asymmetric explosions have in fact been found by Nagataki et al. (\cite{Nagataki_97}) to enhance the production of elements with mass numbers around $A=45$ (i.e., Ca and Ti) and more significantly around $A=65$ (i.e., Zn) (see also Maeda \& Nomoto \cite{Maeda_03}). 

This suggests that stars in the primeval dSph galaxies may have possessed smaller rotation than that of stars belonging to the Milky Way, at least for massive stars. If this is the case, massive dSph stars would produce relatively small amounts of $\alpha$-elements during stellar evolution, and end in spherical SN explosions that produced elements such as Ca, Ti, and Zn less effectively. Figure 4 shows the correlation of [Zn/Fe] with [Mg/Fe] in the dSph galaxies. Essentially, stars with lower [Mg/Fe] exhibit lower [Zn/Fe]. One possible cause of this relation is the end result of the yields affected by the differential stellar rotation; stars exhibiting lower [Mg/Fe] and [Zn/Fe] may be born from the interstellar matter enriched by massive stars with smaller rotation. A large scatter might be attributable to the different production processes between two elements as stated above.

Theoretical calculations have predicted that the metallicity is a key factor to control the stellar rotation in such a way that low-metallicity stars have higher rotational velocities as a result of less angular momentum loss during the main-sequence evolution (Meynet \& Maeder \cite{Meynet_00}; Maeder \& Meynet \cite{Maeder_01}). However, recent direct observations of $V\sin i$ for O-type stars in the Magellanic Clouds (MCs) which have low-metallicities reveal that distributions of rotational velocities for MC stars and Galactic stars are similar, and do not support these predictions (Penny et al. \cite{Penny_04}). As discussed so far, elemental abundances in the LMC also have no signature of the stellar rotation different from Galactic stars.  
Besides, the dSph galaxies of which the metallicity is much lower than those of MCs and the Milky Way are predicted to have stars with smaller rotation in an opposite sense to the metallicity effect. It might suggest that there exists another environmental factor to regulate the stellar rotation. Probably this factor would determine initial rotational velocities resulting from the angular momentum loss during star formation through the interaction between an accretion disk and star by regulating the strength of the magnetic coupling to the disk or the lifetime of accretion disk, and so on.
 
\section{Conclusions}

The [$\alpha$/H] vs.~[Fe/H] diagrams where $\alpha$ represents Mg or average of $\alpha$-elements for stars in the dSph galaxies (i.e., Draco, Ursa Minor, Sextan, Sculptor, Carina, Fornax, and Leo I) present no compelling evidence for the contribution of SNe Ia to stellar abundances in the low-$\alpha$/Fe stars, in contrast with the Sagittarius galaxy and the LMC. It might suggest that massive dwarf galaxies with a mass of several $10^8-10^9$ \ms retain the ejecta of SNe Ia in the gravitational potential well and the interstellar matter in these galaxies is enriched by SNe Ia, while this is not the case for lower mass dSph galaxies. In contrast to such an equivocal fate of the SNe Ia ejecta, a sufficient gas 
should always cool down the ejecta of SNe II and prevent them from escaping from galaxies during star formation, subsequently followed by the galactic wind due to an accumulated SN explosion energy.

The abundance features for other elements in the dSph stars do not support the lack of massive SNe II as well as contributions from SNe Ia as a source of this low-$\alpha$ signature. Rather, the deficiencies in the abundances of $\alpha$-elements and Zn in the dSph stars appear more likely to reflect the heavy-element yields of massive stars with smaller rotation compared to the solar neighborhood stars, although what makes such a difference in stellar rotation is a open question.  Direct observations of $V\sin i$ for stars in the dSph galaxies are needed in order to substantiate these implications. A young population, so-called blue plume, in the Sagittarius galaxy could be a prime candidate for measuring stellar rotation. It should be finally stressed that this hypothesis is speculative and requires further investigation, particularly with respect to the precise effects of rotation on nucleosynthesis for both pre-SN massive stars and SN explosions. The possible environmental factors promoting such rotation will also be studied. 

\begin{acknowledgements}
The author is appreciative of the referee P. Bonifacio for useful  comments that helped improve this paper.
\end{acknowledgements}

\end{document}